# Automated Rain Sampler for Real time pH and Conductivity Measurements

R.M. Weerasinghe[1], A.S. Pannila[1,*], M.K. Jayananda[2] and D.U.J. Sonnadara[2]
[1]*Electro Technology Laboratory, Industrial Technology Institute*
[2]*Department of Physics, University of Colombo, Colombo 3*
*pannila@iti.lk\**

## ABSTRACT

To monitor the acidity of rain water in real time, a rain water sampling system was developed. The rain sampler detects the initial rain after a dry spell and collects a water sample. Before performing the measurements, the pH probe is calibrated using a standard buffer solution whereas the conductivity probe is calibrated using deionized water. After calibrating the probes the pH and the conductivity of the collected rain water sample are measured using the pH and the conductivity probe. Weather parameters such as air temperature, humidity and pressure are also recorded simultaneously. The pH and conductivity measurement data including weather parameters are transmitted to central station using a GSM modem for further analysis. The collected rain water sample is preserved at the remote monitoring station for post chemical analysis. A programmable logic controller controls the entire process.

## 1. INTRODUCTION

At present, hydropower is used as the main power source in Sri Lanka. There are more than eight reservoirs located in the hill country area to store rainfall during the monsoon seasons. The southwest monsoon brings heavy rainfall to the western slopes of the hill country. In contrast, northeastern monsoon season brings rainfall to the north-eastern slopes of the hill country. During the southwest monsoon season the hydro power generation is at maximum whereas during the remaining seasons the hydro power generation is limited [1]. In order to meet the balance power requirement thermal power and renewable power sources are used.

When considering thermal power stations, the main energy source is fossil fuel. Burning fossil fuels produce $SO_2$, $NO_x$, $CO_2$ and other toxic gases. This will cause increase in the carbon foot print. Emission of $SO_2$ and $NO_x$ will increase the acidity of rain water. In addition, the coal fired power plants produce large amounts of fly ash which is ultimately released to atmosphere with the flue gas.

Modern power plants have necessary precautions to collect fly ash from the flue gas to a certain extent. However, being a country located near India where there are several thermal power plants operating in southern parts, Sri Lanka is expected to receive a certain amount of emission during inter monsoon seasons. In order to address the transboundary effects in South Asian countries, the Male Declaration was formed in 1998.

The gasses emitted by burning fossil fuels cause chemical reaction inside the clouds which increase the acidity of rain water. The acidity of rain water depends on the existence of neutralization components like ammonia, calcium carbonate or hydroxide [2]. Normally rain water has a pH value of 5.6 due to dissolved $CO_2$. However gasses from fuel burning further decreases the pH value of rain water. The Central Environmental Authority has set up an acid rain monitoring station at Dutuwewa in north central province under the Male declaration. However, since the pH and conductivity of the samples tend to change with time, the lack of real time measurement in this station makes it difficult to obtain reliable information regarding the acidity of precipitation. Therefore it is necessary to implement an automated system which is capable of monitoring the pH and the conductivity of the initial rain that could precipitate after a long dry spell in real-time.

Several studies have been carried out elsewhere to develop real time acid rain monitoring systems by collecting rain water samples and analyzing the pH and the conductivity. One such study records the results of the tests on a paper media using a printer for later analysis. The collected water samples are preserved for chemical analysis under refrigerated conditions [3]. Studies are available for systems developed using iron chromatograph technology to measure the pH, conductivity and to analyze the $Cl^-$, $NO_3^-$, $SO_4^{2-}$, $Na^+$, $K^+$, $NH_4^+$, $Ca^{2+}$ and $Mg^{2+}$ [4]. Studies are also available on the development of instruments to preserve the water sample, without changing the chemical properties under refrigerated conditions for later analysis [5] as well as on the development of standard reference material for rainfall analysis in an acid rain monitoring system [6].

This work is based on an earlier work carried out to implement an automated weather station [7]. The main objective of this work is to develop a real-time pH and conductivity monitoring system with data communication facility to transmit the results to a central station. The sample is preserved at the remote station for post chemical analysis.

## 2. SYSTEM OVERVIEW

Fig. 1 shows the block diagram of the acid rain monitoring system. The main functions of the system can be divided into five subsystems, the rain water collecting system, the conductivity measurement system, the pH measurement system, the sample preservation system and the data transmission system. The pH measurement system consists of three vessels where two vessels contain deionized water and one vessel with a buffer solution having a pH value of 9.2 ± 0.1. The conductivity measurement system consists of a conductivity probe, a conductivity transmitter and deionized water. Solenoid valves were used to control the flow of liquid to pH and conductivity monitoring chambers. The pH monitoring chamber and the conductivity monitoring chambers were made of glass in order to reduce the contamination of the samples. In an earlier study, a system has been developed based on Z80 microcontroller to control the entire acid rain monitoring station without having a data communication link [3]. In this study the entire system is controlled by a SIEMENS S7 200 PLC. In addition to pH and conductivity measurement system, a separate rain gauge, temperature sensor, humidity sensor and a barometer were interfaced to the PLC in order to measure the weather conditions.

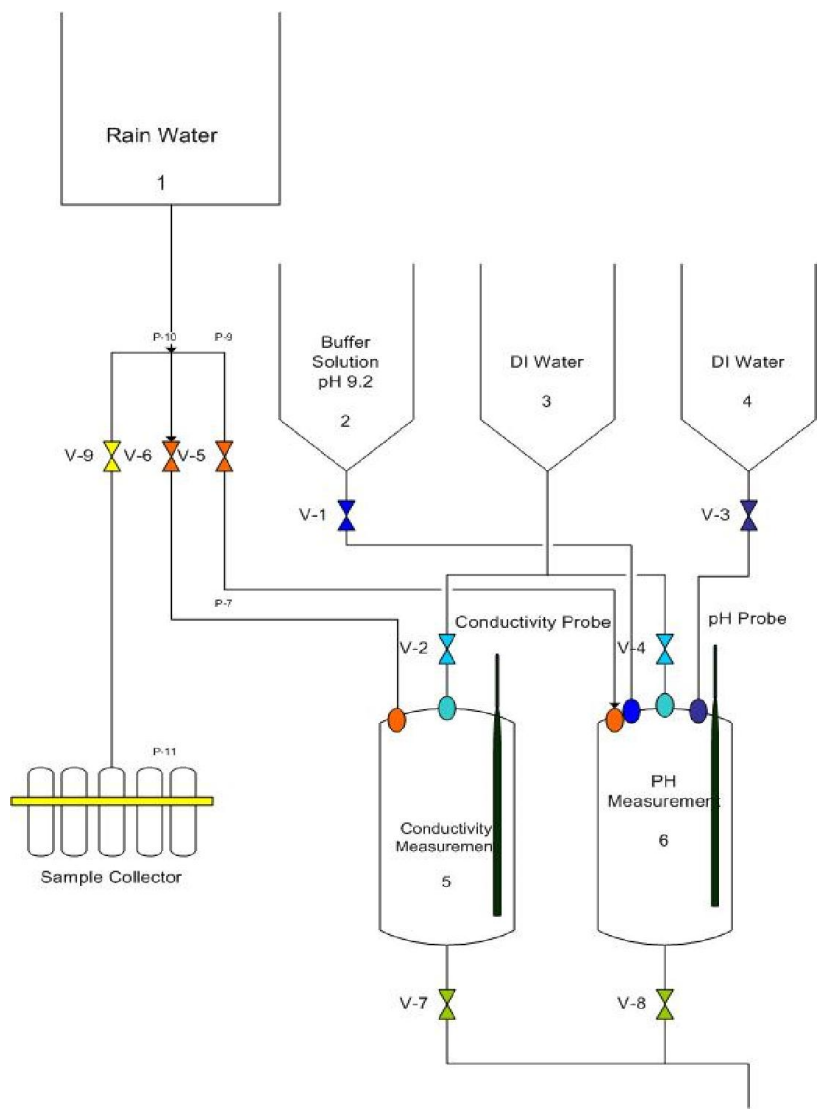

Fig. 1: Diagram of the acid rain monitoring station

## 3. FUNCTIONALITY OF THE SYSTEM

The rain detection sensors are exposed to the atmosphere. When rain drops fall onto two sensors located over the funnel, the cover opens exposing the funnel to the rain. When a predefined amount of rain water (250 ml which is equivalent to 1 mm of rainfall) collected by the rain collector, the cover closes, preventing further collection of rainfall. After collecting the rain water, the rainfall collector goes to a standby mode and the sample analysis process starts.

Normally, the pH probe is immersed in deionized water. Before starting the analysis the deionized water is discharged from the pH measurement chamber by opening the valve V8. The valve is open for 2 s duration. Another valve (V4) opens to allow flow of fresh deionized water to the chamber. The level sensor detects the water level and closes the valve in order to prevent overflows. After 2 s the first valve (V8) opens to remove the deionized water from the chamber. This process repeats three times. The objective of this exercise is to clean the pH probe well before starting the calibration cycle.

After completing the first cleaning cycle, the buffer solution (pH 9.2 solution) is allowed to flow into the chamber by opening the valve V1. When the solution fills to the required level the level sensor detects the liquid level and closes the valve V1. The solution remains inside the chamber for 2 s in order to stabilize the H+ transition with pH probe. The valve V8 opens to discharge the buffer solution from the chamber. The above process repeats three times. Before the end of the third cycle, the pH reading of the solution is recorded through the pH transmitter. The output of the pH transmitter is read by the EM231 ADC module and the value is stored in a two byte memory variable in the PLC.

After completing the calibration process, the pH probe cleaning process starts in order to remove the deposited buffer solution. The valve V4 opens to allow deionized water to flow into the chamber. After 1 s the deionized water is discharged by opening the valve V8. This process continues three times. After completing the cleaning process the rain water analysis process starts.

To allow rain water to flow into the chamber the valve V5 opens. The level sensor detects the rain water level inside the chamber and closes the valve V5. After 1 s the rain water is discharged by opening the valve V8. This process repeats three times. Before discharging the rain water from the last cycle, the ADC module reads the pH value through the pH transmitter. After completing the pH measurement of the rain water, the pH probe is immersed in deionized water. This is done by opening the valve V3. After reaching the correct water level the valve closes automatically.

In order to start the conductivity measurement process, the conductivity probe is calibrated using the deionized water. By opening the valve V2, deionized water is allowed to flow into the conductivity measurement chamber. The valve closes automatically when the liquid level reaches the position of the level sensor. The conductivity value is captured by the ADC through the conductivity probe transmitter. After completing the calibration process, rain water is allowed to flow into the chamber by opening the valve V6. Once the liquid level reaches the level sensor position, the flow of rain water stops automatically. The conductivity reading is captured from the ADC through the pH transmitter after 1 s. Finally by opening the valve V7 the rain water is discharged. The corresponding analog values for the pH, conductivity, temperature and humidity are stored in the memory of the PLC. The PLC uses AT commands to communicate with the GSM modem. The necessary AT commands used to communicate with the PLC is placed in data block while programming the PLC. The data stored in the memory variables are combined into one text message and sent to the modem for transmission.

During the measurement the pH probe checks the pH of a buffer solution in order to assure the accuracy of the pH measurement. The pH probe and the conductivity probe were subjected to a performance test using a standard calibrated pH probe and a conductivity probe in the laboratory. According to the measurements, the pH probe connected to the instrument gives a slight variation with the pH value of the standard probe (laboratory measurement). The variation in the pH reading compared with the standard probe is ±0.2.

Similarly, the variation in the conductivity measurement compared to the laboratory measurement is ±0.5 μS/cm.

## 4. PILOT TEST RESULTS

In order to test the repeatability of the results, two samples of water was prepared having pH values of 4.60 and 5.87 and conductivity values of 460 μS/cm and 64 μS/cm. The two samples were introduced to the precipitation collector at two different measurement cycles. The results are tabulated in Table 1. During the pilot test, the temperature and the humidity inside the laboratory were maintained at fixed values.

Table 1: Results of the laboratory measurements

| Sample No | Cycle No | pH | Conductivity (μS/cm) |
|---|---|---|---|
| 1 | 1 | 4.65 | 446 |
|   | 2 | 4.66 | 448 |
| 2 | 1 | 5.80 | 64.7 |
|   | 2 | 5.62 | 65 |

Table 2: Results of the field measurements

| Sample no | pH | Conductivity |
|---|---|---|
| 1 | 7.08 | 11.79 |
| 2 | 6.59 | 86.79 |
| 3 | 6.66 | 93.04 |
| 4 | 7.03 | 57.62 |
| 5 | 6.49 | 63.87 |
| 6 | 6.38 | 60.65 |
| 7 | 6.67 | 65.95 |
| 8 | 6.52 | 18.04 |
| 9 | 6.00 | 11.79 |
| 10 | 6.16 | 11.79 |

According to the results given in Table 1, the cycle 1 and cycle 2 give similar results for the sample no 1 and sample no 2 within the estimated errors for pH and conductivity. This test confirms the repeatability of the results.

In order to check the performance of the instrument with real rainfall, the instrument was installed within the premises of the Industrial Technology Institute. The location where the instrument was installed is an open space. The pH and conductivity data shown in Table 2 were obtained for ten separate rainfall samples. According to the collected data acidity was not detected in the collected rain water samples.

## 5. CONCLUSIONS

In this work, an instrument to detect acidity levels of rainfall was successfully developed and tested. Laboratory and pilot tests indicated that the instrument worked as expected. However it is necessary to place the instrument in remote locations to monitor the acidity of rainfall. The work is in progress to simulate the path of propagation of possible pollutants emitted from coal power plants to identify suitable locations.